\documentclass[fp,twocolumn]{jpsj3}

\usepackage{txfonts}

\usepackage{amsmath}
\usepackage{amsfonts}
\usepackage{amssymb}
\usepackage{graphicx}
\usepackage[dvipsnames]{xcolor}
\usepackage{tikz}

\usepackage{bm}

\usepackage{mathtools}

\title{Block-Lanczos Density-Matrix Renormalization-Group Approach to Spin Transport in Heisenberg Chains Coupled to Leads}

\author{Florian Lange$^1$, Satoshi Ejima$^{1,3}$, Tomonori Shirakawa$^2$, Seiji Yunoki$^{2,3,4}$, and Holger Fehske$^1$}

\inst{$^1$Institut f\"ur Physik, Universit\"at
Greifswald, 17489 Greifswald, Germany \\
$^2$Computational Materials Science Research Team, RIKEN Center for Computational Science (R-CCS), Kobe, Hyogo 650-0047, Japan \\
$^3$Computational Condensed Matter Physics Laboratory, RIKEN Cluster for Pioneering Research (CPR), Wako, Saitama 351-0198, Japan \\
$^4$Computational Quantum Matter Research Team, RIKEN Center for Emergent Matter Science (CEMS), Wako, Saitama 351-0198, Japan}

\abst{
We adapt the block-Lanczos density-matrix renormalization-group technique to study the spin transport in a 
spin chain coupled to two non-interacting fermionic leads. As an example, we consider leads described by two-dimensional tight-binding 
models on a square lattice. 
Although the simulations are carried out using a chain representation of the leads, observables in the original two-dimensional lattice can be calculated by reversing the block-Lanczos transformation. 
This is demonstrated for leads with Rashba spin-orbit coupling. 
}

\begin{document}

\maketitle

\section{Introduction}
Magnetic insulators are potentially useful for spintronics applications because of the greater decay length of spin currents compared to conductors. 
In this light, there have been several experiments on the spin-current generation in insulating materials.\cite{SpinWaveSpinCurrent,Uchida2010,SpinWaveSpinCurrentExperiment,SpinonSpinCurrent} 
For example, Kajiwara \textit{et al.} achieved the injection of spin currents into the ferrimagnetic insulator yttrium iron garnet (YIG) from an attached platinum (Pt) film, where a spin current was generated electrically using the spin Hall effect.\cite{SpinWaveSpinCurrent} 
Spin currents have also been experimentally investigated for other types of magnets, including various antiferromagnets~\cite{AFMSpintronics}, the paramagnet Gd$_3$Ga$_5$O$_{12}$\cite{SpinCurrentParamagnet} and the spin nematic liquid LiCuVO$_4$\cite{SpinCurrentNematic}. 
Relevant to the present study, a spin current was induced in the spin-1/2-chain material Sr${}_2$CuO${}_3$ by using the longitudinal spin Seebeck effect\cite{SpinSeebeckFirst} at an interface with a Pt film.\cite{SpinonSpinCurrent} 
In the same work, the spin transport for this setup has been studied theoretically using a Green's function technique.\cite{PhysRevB.50.5528} See also Refs.~\citen{SpinTransportLinearResponseFM,SpinCurrentNematic} for similar analyses of spin transport in other magnetic systems.

Motivated by these developments, we numerically study the spin transport in a one-dimensional (1D) antiferromagnetic spin-1/2 chain coupled to two non-interacting fermionic leads. In numerical simulations, the leads are typically represented by a finite number of sites which causes a discretization of the Hamiltonian in energy space. 
To reach a larger number of sites and a finer discretization of energy, it is advantageous to map the non-interacting leads to 
a chain representation that permits the use of matrix-product state (MPS) techniques, e.g., the density-matrix renormalization-group (DMRG) or the numerical renormalization group method.\cite{White92,Sch11} 
There are different ways to obtain such a chain mapping. 
Here, we use a Lanczos technique,\cite{BlockLanczosDMRG} which is convenient when starting from a model defined on a real-space lattice.

Previously, we investigated the spin transport for a junction in which the leads 
are modelled by tight-binding chains with uniform hopping parameters.\cite{Lange_2018,Lange_2019} 
The present study differs in that we start from a higher-dimensional tight-binding model and obtain the chain representation numerically. 
This serves two purposes: (i) we can evaluate how important the assumption of homogeneous 1D leads is for the conclusions of Refs.~\citen{Lange_2018,Lange_2019}, (ii) we outline a general strategy for the numerical study of spin transport in 1D systems coupled to non-interacting leads of arbitrary dimension. 
Here, we specifically consider a junction with two-dimensional (2D) leads defined on a square lattice. 
Within the 1D representation of this model, we calculate the spin conductance with the DMRG and the time-evolving block decimation methods,~\cite{PhysRevLett.91.147902} finding qualitatively similar results to the setup with uniform 1D leads. 
Observables in the original 2D lattice can also be calculated by reversing the block-Lanczos transformation. 
We demonstrate this for leads with Rashba spin-orbit coupling, which is important for semiconductor heterostructures.\cite{Manchon2015}
Note that although we focus on a specific type of spin chain and 2D leads, the presented numerical approach could be applied to other models, e.g., different types of spin chains or other lattices for the leads. 

The rest of this paper is organized as follows. 
In Sec.~\ref{secmodel}, we introduce the model. 
Section~\ref{secmethod} describes the block-Lanczos transformation applied to 
facilitate the use of MPS methods. 
The results of the MPS calculations are presented in Sec.~\ref{secres}. 
Section~\ref{secsummary} contains a summary and a discussion of the results.

\section{Model}
\label{secmodel}

\begin{figure}[b]
\centering
\includegraphics[width=0.85\linewidth]{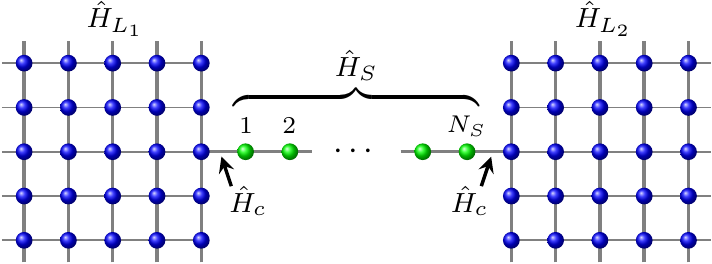}
\caption{(Color online) Schematic of the junction with two Rashba leads. The green spheres represent spin-chain sites, the blue ones lead sites. 
}
\label{figsys}
\end{figure}

We consider a junction of a 1D interacting region and two non-interacting leads (see Fig.~\ref{figsys}). 
The interacting part is modeled by a spin-1/2 Heisenberg chain of length $N_S$,
\begin{align}
\hat{H}_S = J \sum_{j=1}^{N_S-1} \bm{\hat{S}}_j \bm{\hat{S}}_{j+1} \, ,
\end{align}
with 
antiferromagnetic exchange coupling $(J > 0)$. 
For the lead Hamiltonians $\hat{H}_{L_1}$ and $\hat{H}_{L_2}$, we assume 
2D tight-binding models, possibly with additional Rashba spin-orbit coupling, i.e.,
\begin{align}
\hat{H}_{L_1(L_2)} = \sum_{\langle i, j \rangle} &\left(\hat{c}_{i\uparrow}^{\dagger},\hat{c}_{i\downarrow}^{\dagger}\right) \{ -t -{\rm i}\lambda [(x_j-x_i) \sigma_y     
 \nonumber \\[-3mm] -&(y_j-y_i) \sigma_x] \} \begin{pmatrix}\hat{c}_{j\uparrow} \\ \hat{c}_{j\downarrow} \end{pmatrix} 
 -\mu\sum_{i,\sigma}\hat{c}_{i\sigma}^\dag\hat{c}_{i\sigma}^{\phantom{\dagger}}\, ,
\label{eqRashbaLead}
\end{align}
where $\hat{c}_{j\sigma}^{(\dagger)}$ is the annihilation (creation) operator of an electron with spin $\sigma \in \{\uparrow,\downarrow\}$ 
at site $j$ [located at $(x_j,y_j)$] on an infinite square lattice with an open edge and lattice constant $a=1$, and 
$\langle i, j \rangle$ indicates all pairs of nearest-neighbor sites $i$ and $j$. 
The spin-orbit coupling is parametrized by $\lambda$, $\sigma_x$ and $\sigma_y$ are Pauli matrices, 
and $\mu$ is the chemical potential. 
At each end, the spin chain is coupled via an exchange interaction to a site at the open edge of 
one of the leads. 
Denoting the indices of these sites by $j_0$, the coupling terms are of the form 
\begin{align}
\hat{H}_c(a,j) &=  J' \left[ \frac{1}{2} (\hat{c}_{j_0\uparrow a}^{\dagger} \hat{c}_{j_0\uparrow a}^{\phantom{\dagger}} - \hat{c}_{j_0\downarrow a}^{\dagger} \hat{c}_{j_0\downarrow a}^{\phantom{\dagger}}) \hat{S}_j^z \right. \nonumber \\ &  \hspace*{1.cm} + \left. \frac{1}{2} (\hat{c}_{j_0\uparrow a}^{\dagger} \hat{c}_{j_0\downarrow a}^{\phantom{\dagger}} \hat{S}_j^- + 
\hat{c}_{j_0\downarrow a}^{\dagger} \hat{c}_{j_0\uparrow a}^{\phantom{\dagger}} \hat{S}_j^+) \right] \, ,
\label{eqcoupling}
\end{align}
where an additional index $a \in \{L_1,L_2\}$ was introduced to distinguish between the two leads. 
With Eq.~\eqref{eqcoupling}, the complete Hamiltonian takes the form 
\begin{align}
\hat{H} &= \hat{H}_S + \hat{H}_{L_1} + \hat{H}_{L_2} + \hat{H}_c(L_1,1) + \hat{H}_c(L_2,N_S) \,. 
\end{align}

\section{Block-Lanczos Transformation}
\label{secmethod}

The Lanczos algorithm is a way to obtain a unitary transformation that tridiagonalizes a given Hermitian matrix $H^0$.\cite{SaadIterMethods} 
One starts with a single unit vector $v_1$ which is the first column of the transformation matrix $P=(v_1,v_2,...)$. 
All remaining $v_j$ are then obtained by setting $v_j \leftarrow H^0 v_{j-1}$ ($j=2,3, 4, \dots$) 
and orthogonalizing against previous vectors.  
We use a block version of the Lanczos method,\cite{BlockLanczosRuhe} in which one chooses the first $M$ orthonormal vectors 
$v_1,v_2, \dots,v_M$ and sets 
$v_n \leftarrow H^0 v_{n-M}$, 
again followed by an orthogonalization.  
From this construction and the Hermicity of $H^0$, it follows that $P^{\dagger} H^0 P$ is a band matrix with bandwidth $2M+1$: 
\begin{align}
 \tilde{H}^{0} = P^{\dagger} H^{0} P &= \left(
\begin{array}{ccccc}
E_1 & T_1 & 0 & 0 & \cdots \\
T_1^{\dagger} & E_2 & T_2 & 0 & \cdots \\
0 & T_2^{\dagger} & E_3 & T_3 &  \\[-2mm]
0 & 0 & T_3^{\dagger} & E_4 & \ddots  \\
\vdots & \vdots &  &  \ddots &  \ddots
\end{array}\right) \, ,
\end{align}
where $E_j$ and $T_j$ are Hermitian and lower-triangular $M\times M$ matrices, respectively.

If the block-Lanczos transformation is applied to the matrix $H^0$ describing a single-particle Hamiltonian $\hat{H}^0$ of 
a system $\hat{H}^0 = \bm{\hat{c}}^{\dagger} H^0 \bm{\hat{c}}$, 
where $\bm{\hat{c}}^{\dagger} = (\hat{c}_1^{\dagger},\hat{c}_2^{\dagger},...,\hat{c}_N^{\dagger})$ is a vector of creation operators 
(here the spin index is implicitly assumed), 
the banded structure of $\tilde{H}^0$ means that, in terms of the new operators $\bm{\hat{a}}^{\dagger} = (\hat{a}_1^{\dagger},\hat{a}_2^{\dagger},...,\hat{a}_N^{\dagger}) = \bm{\hat{c}}^{\dagger} P $, $\hat{H}^0$ describes an open chain with short-ranged hopping and a site-dependent potential. 
Explicitly stated:
\begin{align}
\hat{H}^0 =  \bm{\hat{a}}^{\dagger} \tilde{H}^0 \bm{\hat{a}} = &\sum_{d=1}^{M} \sum_{n=1}^{N-d}  \left( \tilde{H}_{n,n+d}^0 \hat{a}_n^{\dagger}\hat{a}_{n+d}^{\phantom{\dagger}} + \rm{h. \, c.} \right) \nonumber \\  &+ \sum_{n=1}^{N} \tilde{H}_{n,n}^0 \hat{a}_n^{\dagger}\hat{a}_{n}^{\phantom{\dagger}} \, ,
\end{align}
where $N$ is the order of the matrix $H^0$, corresponding to the number of single-particle states. 
The first $M$ operators $\hat{a}_1^{\dagger},\hat{a}_2^{\dagger},...,\hat{a}_M^{\dagger}$ 
can be fixed through the choice of initial vectors.
Increasing $M$ thus allows for a greater flexibility in the  transformation but it also increases the maximum hopping range.

For an interacting Hamiltonian $\hat{H} = \hat{H}^0 + \hat{V}$, in which the interaction is a function of only $M$ electron operators, i.e., 
$\hat{V} = \hat{V}(\hat{c}_{j_1}^{\dagger},...,\hat{c}_{j_M}^{\dagger},\hat{c}_{j_1},...,\hat{c}_{j_M})$, we can use a block-Lanczos transformation with block size $M$ and set $\hat{a}_n^{\dagger}= \hat{c}_{j_n}^{\dagger}$ for $n=1,2,...,M$. 
The interaction is then restricted to the first $M$ sites after the transformation and the chain representation of $\hat{H}$ 
 contains only short-ranged terms. 
An exact solution of the problem will still be impossible in general but the 1D form makes the model suitable 
for a numerical treatment 
with MPS techniques. 
The Lanczos method has been used in this manner mostly in the context of impurity problems.\cite{BlockLanczosDMRG,LanczosDMRG} 
Our application here to the spin-chain junction is similar to it, with the sites coupled to the spin chain taking the role of the impurities.

\subsection{Infinite boundary conditions}
\label{secmethodIBC}
In the following discussion, we assume that the Hamiltonian is originally defined on a lattice in real space and has only short-ranged 
hopping terms. The interaction $\hat{V}$ shall act on one site whose corresponding fermion operators (with internal degrees of 
freedom, such as spin, and thus $M$ being larger than 1)
will be invariant under the transformation. 
While we are ultimately interested in the thermodynamic limit, the matrix $H^0$ and thus the system size need to be finite in a 
numerical calculation. 
However, for a given $j_{\rm max}$ one can always choose the original system large enough so that the transformed operators 
$\hat{a}_j^{\dagger}$ with $j \leq j_{\rm max}$ are not affected by its finite size. 
This follows from 
\begin{align}
v_j &\in \text{span}\left\{ (H^0)^n v_l \ | \ 1 \leq l \leq M, \, 0 \leq n \leq r-1 \right\} 
\end{align}
for $j \leq rM$, which implies that  
$\hat{a}_j^{\dagger}$ is supported only on sites connected to the interacting site through at most $r-1$ hopping operations. 
Of course, finite-size effects will eventually appear if the transformation is carried out to completion. 
We can, however, stop the Lanczos recursion before that happens and work with a truncated transformation by ignoring the remaining sites in the chain representation.  
For a fixed number of chain sites these {\it infinite boundary conditions} will be closer to the thermodynamic limit 
with regard to the physics at the interacting site.\cite{LanczosDMRG} 

On the downside of this procedure, 
$P$ is no longer unitary which complicates the measurements for the original lattice. 
The one-body expectation values in both representations are related by
\begin{equation}
\langle \hat{c}_j^{\dagger} \hat{c}_i^{\phantom{\dagger}} \rangle = \sum_{n,m} P_{mj}^{\dagger} P_{in}^{\phantom{\dagger}} \langle \hat{a}_m^{\dagger} \hat{a}_n^{\phantom{\dagger}} \rangle \, 
\label{eqExpValBL}
\end{equation}
for the full transformation. 
Typical quantities in the original system thus have to be reconstructed from a large number of correlation functions in the effective 1D model. 
If the Lanczos recursion is stopped prematurely, 
the above relation is not fulfilled because states are missing on the right-hand side. 
However, it is still possible to calculate with reasonable accuracy the change in the expectation values that is induced by a perturbation 
at the interacting site, e.g., an injected current. 
The reason is that, for sufficiently many sites in the effective 1D model, the only missing terms in Eq.~\eqref{eqExpValBL} are then 
between sites that are both outside the range of the perturbation, or whose distance so large that the contribution from the correlation 
function can be neglected. 
We thus expect the following relation to hold during the time evolution:
\begin{align}
\langle \hat{c}_j^{\dagger} \hat{c}_i^{\phantom{\dagger}} \rangle(\tau) - \langle \hat{c}_j^{\dagger} \hat{c}_i^{\phantom{\dagger}} \rangle(0) & \approx {\sum_{n,m}}' P_{mj}^{\dagger} P_{in}^{\phantom{\dagger}} [ \langle \hat{a}_m^{\dagger} \hat{a}_n^{\phantom{\dagger}} \rangle(\tau)  \nonumber \\ &\hspace*{2cm} - \langle \hat{a}_m^{\dagger} \hat{a}_n^{\phantom{\dagger}} \rangle(0)] \, ,
\label{eqExpValBLapprox}
\end{align}
where $\tau$ denotes the time at which the expectation value is calculated and $\sum'$ indicates that the sum over $m$ and $n$ 
is truncated. 

Note that another more accurate way to calculate expectation values away from the interacting site 
is to increase the block size $M$ and include additional sites in the initial block-Lanczos basis set for measurements. 
This has been used in Ref.~\citen{BlockLanczosDMRG} to calculate two-point correlation functions between impurity and conduction sites. 
For our problem this scheme is less suitable, since we intend to carry out measurements at a lot of different positions, 
which would require the simulations to be repeated many times.

\subsection{Application to leads}

\begin{figure}[tb]
\centering
\includegraphics[width=0.65\linewidth]{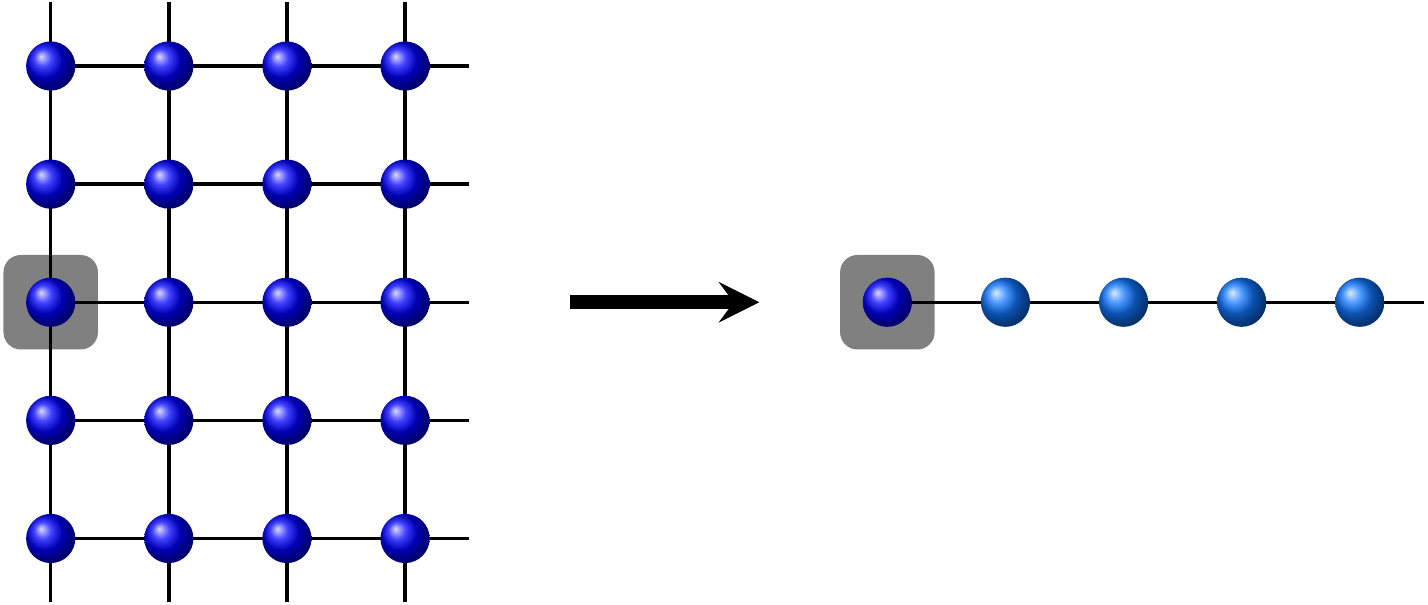}
\caption{(Color online) Block-Lanczos transformation mapping the 2D Rashba lead (spin degrees of freedom are implicitly assumed) 
to a chain with nearest-neighbor hopping. 
Since the only site affected by the interaction 
with the spin chain (shaded regions) is invariant under the transformation, the interaction remains local in the quasi-1D representation. 
}
\label{figbl}
\end{figure} 

We now apply the block-Lanczos transformation to a 2D tight-binding lead with Rashba spin-orbit coupling described in Sec.~\ref{secmodel}. 
To keep the exchange interaction with the spin chain local, both up and down spin states of the affected site on the lead are included 
in the initial block-Lanczos basis set, and the block size is therefore  $M=2$. 
Different transformations and the resulting effective 1D models are obtained, depending on the position of the interacting site. 
Here, the site is assumed to lie on the open edge of the lead (see Fig.~\ref{figbl}).

\begin{figure}[tb]
\centering
\includegraphics[width=0.86\linewidth]{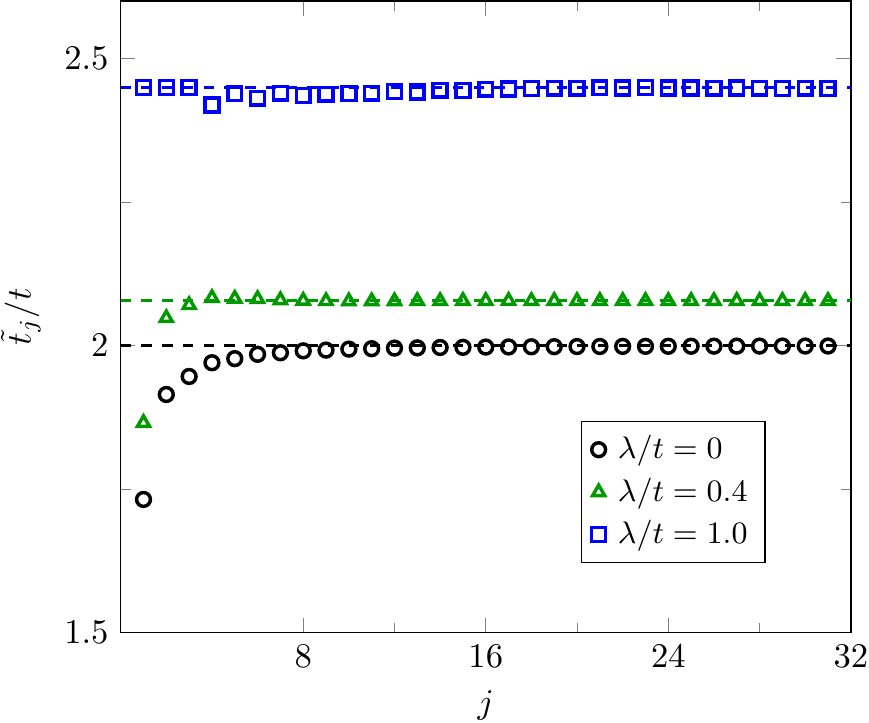}
\caption{(Color online) Hopping parameters $\tilde{t}_j$ in the effective 1D model for the Rashba Hamiltonian $\hat{H}_{L_1(L_2)}$
described by Eq.~\eqref{eqRashbaLead}. The spin-orbit interaction changes the asymptotic value $\tilde{t}_{\infty}$ (dashed lines) 
given by Eq.~\eqref{eqtinf} and slightly affects the position dependence for small $j$. }
\label{figtj}
\end{figure}

It turns out that $\hat{H}_{L_1(L_2)}$ expressed in terms of the new fermion operators $\hat{a}_{j}^{\dagger}$ describes two decoupled tight-binding chains with the same bond-dependent hopping parameters $\tilde{t}_j$. 
Defining $\hat{a}_{2j-1}\to\hat{a}_{j\uparrow}$ and $\hat{a}_{2j}\to\hat{a}_{j\downarrow} $ allows us to write 
\begin{align}
\hat{H}_{L_1(L_2)} &= \sum_{j\geq 1} \tilde{t}_j \sum_{\mathclap{\sigma=\uparrow,\downarrow}}  \left( \hat{a}_{j\sigma}^{\dagger} \hat{a}_{j+1 \sigma}^{\phantom{\dagger}} + \ {\rm h.c.} \right) 
-\mu\sum_{j,\sigma}\hat{a}^\dag_{j\sigma}\hat{a}_{j\sigma} \, .
\label{eqRashbaLeadTransf}
\end{align} 
Therefore, even for the finite Rashba spin-orbit interaction the only difference to a regular 
tight-binding chain is the position-dependent hopping $\tilde{t}_j$. 
The chemical potential term remains the same under the block-Lanczos transformation 
but the density changes in general unless the system is at half-filling. 
The conservation of the new spin introduced in Eq.~\eqref{eqRashbaLeadTransf} can be exploited in numerical simulations.  
Note, however, that it corresponds to the physical spin only at the first site.

The simple form of Eq.~\eqref{eqRashbaLeadTransf} for the Hamiltonian in the block-Lanczos basis implies that the Krylov subspaces 
generated by the $| {\uparrow} \rangle = \hat{a}_{1\uparrow}^{\dagger} | {0} \rangle$ single-particle state at the interacting site are orthogonal to those generated by the $|{\downarrow} \rangle  = \hat{a}_{1\downarrow}^{\dagger} | {0} \rangle$ state. 
To prove this, one needs to show that 
$\langle {\uparrow} | \hat{H}_{L_1(L_2)}^n | {\downarrow} \rangle = 0$
for all natural numbers $n$. 
This follows from the time-reversal symmetry of the Hamiltonian: 
\begin{align}
\langle {\uparrow} | \hat{H}_{L_1(L_2)}^n | {\downarrow} \rangle &
= \langle \tilde{\downarrow} | \hat{T} \hat{H}_{L_1(L_2)}^n \hat{T}^{-1} | \tilde{\uparrow} \rangle 
= - \langle {\uparrow} | \hat{H}_{L_1(L_2)}^n | {\downarrow} \rangle\, ,
\end{align}
where $|\tilde{\downarrow} \rangle = \hat{T} | {\downarrow} \rangle = -|{\uparrow }\rangle$, $|\tilde{\uparrow} \rangle = \hat{T} | {\uparrow} \rangle = |{\downarrow} \rangle$, 
and $\hat{T}$ is the time-reversal operator, assuming that $t$ is real in Eq.~(\ref{eqRashbaLead}). 
The Lanczos method with block size $M=1$ therefore would have been sufficient to obtain the transformation. 
This is also true if there is both Rashba and Dresselhaus spin-orbit coupling, but not in the presence of a magnetic field.

Figure~\ref{figtj} shows the position dependence of the hopping amplitudes in Eq.~\eqref{eqRashbaLeadTransf} for the infinite boundary conditions described in the previous section. Already after a few sites $(j\approx 8)$, the hopping amplitude approaches a constant value that depends on the spin-orbit interaction $\lambda$. 
This asymptotic value agrees with the hopping 
\begin{align}
\tilde{t}_{\infty} &= 2t\cos(k_m) + \sqrt{2} \lambda \sin(k_m) 
\label{eqtinf}
\end{align}
with $k_m=\arctan(\lambda/\sqrt{2}t)$, 
which leads to the same bandwidth as in the original Hamiltonian in Eq.~(\ref{eqRashbaLead}). 
Note that if we applied the transformation to a finite and smaller system, 
the hopping parameter would become non-uniform for sufficiently larger site indices.

From Eq.~\eqref{eqRashbaLeadTransf} and Fig.~\ref{figtj}, one can see that, regardless of the spin-orbit coupling strength 
$\lambda$, the effective model is quite similar to a regular tight-binding chain with essentially uniform hopping amplitude 
except for the first few sites. 
The differences due to the Rashba spin-orbit coupling only become apparent when transforming back to the original representation.

\section{Matrix-product-state calculations}

\label{secres}
\subsection{Spin conductance}

With the 2D Rashba system mapped to a chain representation, we are now in the position to examine the spin transport in the junction  
by means of MPS techniques. 
Because of the Rashba spin-orbit coupling in the leads, we could induce a spin current by exploiting the spin Hall effect and applying an electric field. 
However, to simulate an electric field in the Rashba system, we need to either switch on a static potential at the start of the time evolution or add a time-dependent phase factor to the hopping terms. 
In both cases, the perturbation would be highly non-local in the block-Lanczos basis, rendering MPS simulations inefficient. 
We therefore neglect the Rashba spin-orbit coupling in this section and assume that the spin current is driven by an effective spin-voltage.  
Namely, we add a potential term $\hat{H}_V = (V/2)\sum_{j \in L_1}(\hat{c}_{j\uparrow}^{\dagger}\hat{c}_{j\uparrow}^{\phantom{\dagger}} - \hat{c}_{j\downarrow}^{\dagger}\hat{c}_{j\downarrow}^{\phantom{\dagger}})$ to the first lead, which induces a spin current polarized in the $z$ direction.

The spin conductance $G_S=I/V$ is defined as the ratio of the spin current $I$ that flows through the junction in the nonequilibrium steady state, and the spin voltage $V$. 
For the steady-state spin current, one can write $I = \lim_{\tau \rightarrow \infty} \langle \hat{j}_\ell^z(\tau) \rangle$ ($1 \leq \ell < N_S$), where $\tau$ is the time and $\hat{j}_\ell^z= {\rm i}(J/2)(\hat{S}_\ell^+\hat{S}_{\ell+1}^- - \hat{S}_\ell^-\hat{S}_{\ell+1}^+)$ the spin-current operator between sites $\ell$ and $\ell+1$ of the spin chain. 
To obtain the zero-temperature spin conductance $G_S$ of the junction numerically, we first calculate the ground state with the DMRG method 
and then simulate the time evolution with switched-on spin voltage using the time-evolving block decimation 
algorithm. 
Because of the finite lead sizes, a true steady state is not reached in the simulations. However, it is nevertheless possible to estimate $I$ accurately from the time-dependence of the local spin currents $\langle\hat{j}_\ell^z(\tau)\rangle$. 

More details on the numerical method are given in Refs.~\citen{Lange_2018,Lange_2019}, 
where a similar setup with uniform 1D leads was studied. 
In these works, it was found that the spin conductance $G_S$ 
depends sensitively on the model parameters near the interfaces.  

Generally, $G_S$ is significantly reduced compared with the spin conductance $G_S^0=1/(4\pi)$ for a homogeneous tight-binding chain, i.e., a system without spin chain.  
Through fine-tuning of the parameters at the interfaces, however, a so-called conducting fixed point may be reached.\cite{ConductingFixedPoint1,ConductingFixedPoint3}   
There, the zero-temperature spin conductance is not reduced by interface effects and takes the maximum value $G_S^0$ determined by the leads.\cite{MaslovStone,SafiSchulz} 
Using the specific 2D leads (in their 1D representation) should not qualitatively affect this result but the 
site dependence of the hopping parameters near the interface could move the system away from or towards a conducting fixed point. 

\begin{figure}[tb]
\centering
\includegraphics[width=0.85\linewidth]{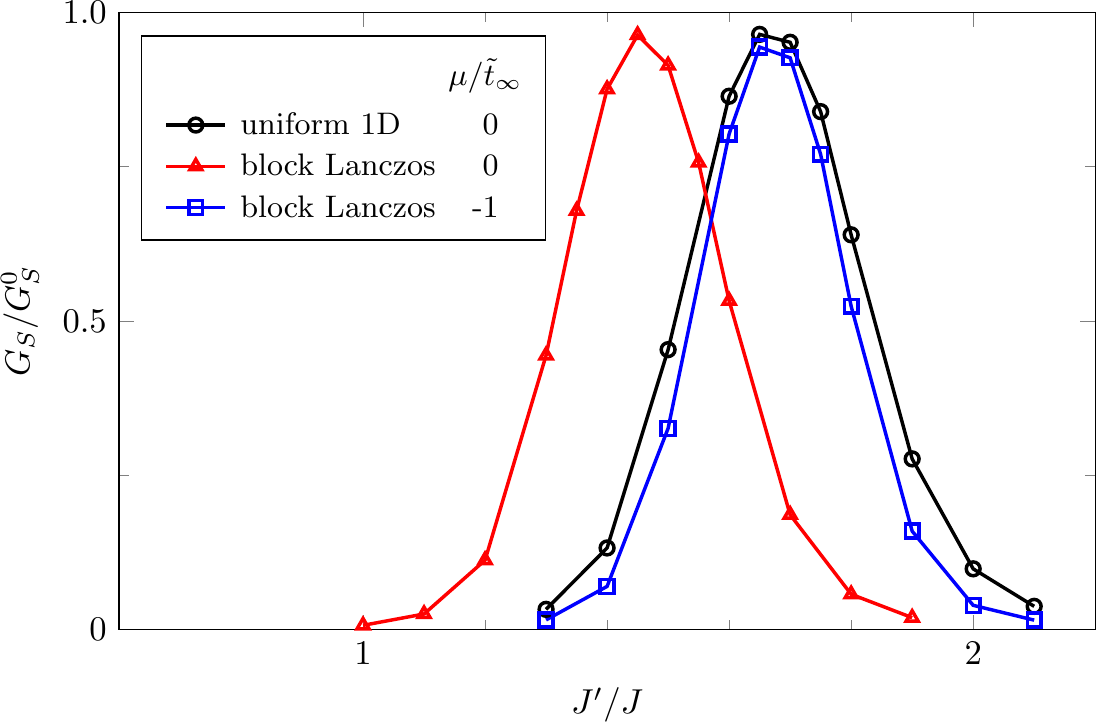}
\caption{(Color online) Spin conductance $G_S$ at zero temperature for $N_S=12$ and spin voltage $V_S/J = 0.1$. 
The hopping amplitudes $\tilde{t}_j$ [see Eq.~\eqref{eqRashbaLeadTransf}] in the leads are either assumed to be uniform or obtained 
by the block-Lanczos transformation of $\hat{H}_{L_1(L_2)}$ in Eq.~\eqref{eqRashbaLead} without spin-orbit coupling $(\lambda=0)$. 
In both cases, the overall energy scale is chosen so that $\tilde{t}_{\infty}=\lim_{j\rightarrow \infty} \tilde{t}_j = J$. 
Each lead is truncated to a finite length of $400$ sites (without including spin degrees of freedom)
in the MPS simulations. 
}
\label{figGs}
\end{figure}

We have confirmed this for some values of the model parameters by explicit numerical calculations, as shown in Fig.~\ref{figGs}. 
At a small but finite spin voltage $V_S/J=0.1$, the spin conductance $G_S$ shows a sharp peak as a function of the interface-coupling strength $J'$. 
This is observed for both types of leads but the position of the maximum is different. For the block-Lanczos leads, the peak is shifted to smaller $J'$, as may be expected because of the reduced hopping strength near the spin chain. 
The maximum value of $G_S$ is in both cases approximately the ideal value $G_S^0=1/(4\pi)$ for the linear conductance. 
In Refs.~\citen{Lange_2018,Lange_2019}, only half-filled leads were considered. As shown in Fig.~\ref{figGs}, 
conducting fixed points occur for a finite chemical potential $\mu$ as well, 
although their position (i.e., $J'/J$) is $\mu$ dependent.

It should be noted that the coupling strength $J_c'$ between spin chain and metallic leads corresponding to the conducting fixed point in Fig.~\ref{figGs} is larger than the exchange coupling $J$ inside the spin chain. In a real experiment, on the other hand, $J_c'$ may be much smaller than $J$, so that 
according to our simple model the system would be far away from the conducting fixed point. However, our calculations were done for the zero temperature limit in which the reduction of the conductance away from the conducting fixed point is particularly strong.\cite{ConductingFixedPoint1,ConductingFixedPoint3} 
An interesting open question is, how the interface effects change quantitatively when a more realistic description is used that, e.g., also takes finite temperature into account.

Finally, let us briefly comment on the case of ferromagnetic exchange interaction $J$ and $J'$. 
Numerical calculations indicate that 
the spin currents are much smaller than for the antiferromagnetic spin-1/2 chains studied here. 
This may be explained by considering the Kondo model, i.e., a single spin coupled to a fermionic bath. 
It is known that for ferromagnetic interaction, the Kondo spin effectively decouples from the lead in the low temperature limit.\cite{PoorMansScalingKondo} Adding spin sites that interact ferromagnetically with the first spin will not change this behavior. 
Accordingly, we expect the linear spin conductance of the junction to vanish at zero temperature in the case of ferromagnetic 
exchange interaction.

\subsection{Lead dynamics}
\label{secreslead}
In the previous section, we only considered expectation values in the effective 1D model, which is sufficient to characterize the spin transport through the spin chain. 
As described in Sec.~\ref{secmethodIBC}, however, the block-Lanczos transformation can be reversed to obtain expectation values of observables 
defined on the original real-space lattice. 
To this end, we calculate the single-particle expectation values $\langle \hat{a}_{i\sigma}^{\dagger} \hat{a}_{j\sigma}^{\phantom{\dagger}} \rangle$ 
in the 1D representation of the lead for all sites $i,j \leq L$ with some finite number $L$. 
This can be done by two nested sweeps in the MPS calculation so that the computational cost scales quadratically with $L$. 

\begin{figure}[tb]
\centering
\includegraphics[width=0.96\linewidth]{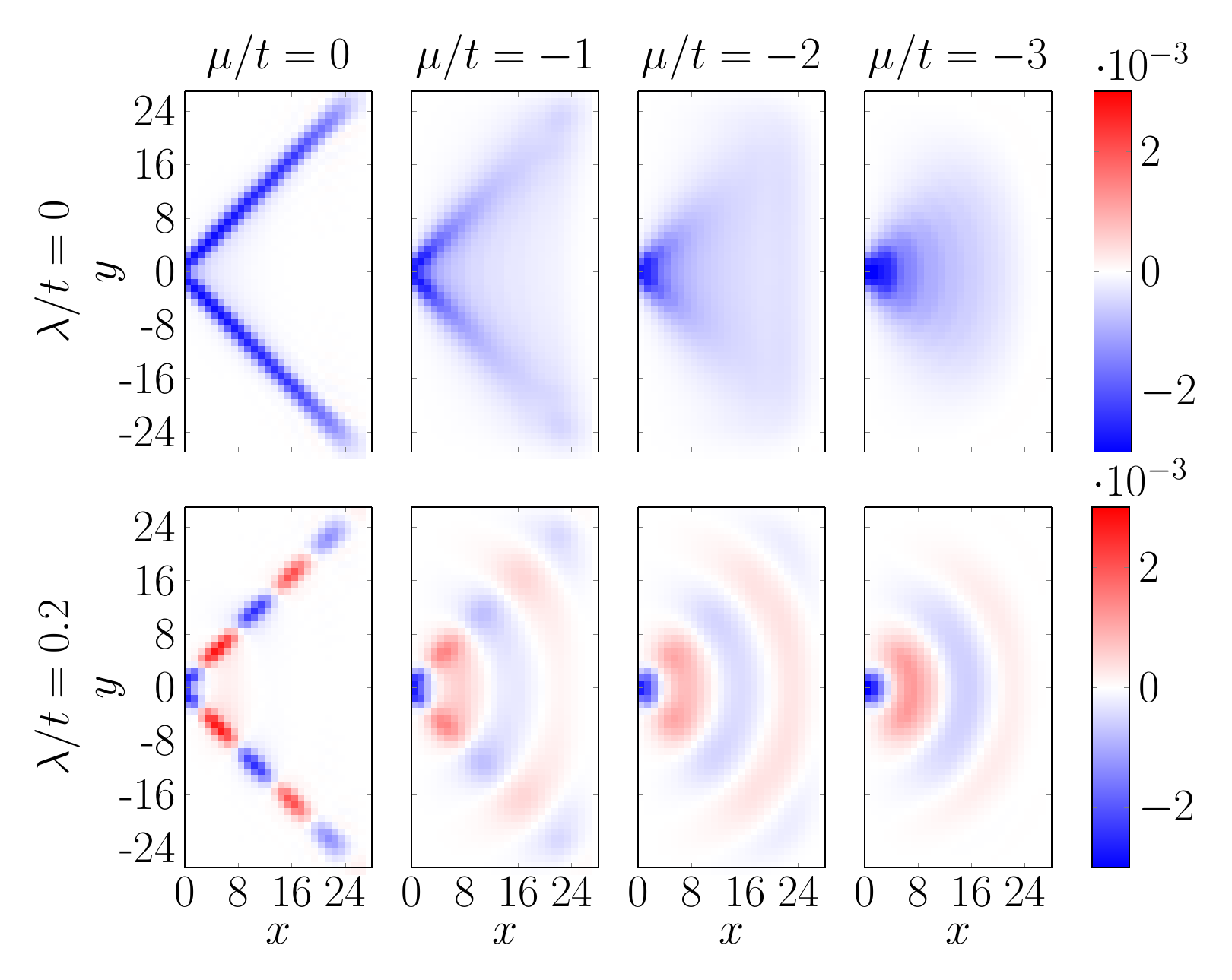}
\caption{(Color online) 
Local magnetization $m_i = \frac{1}{2} \langle \hat{c}_{i\uparrow}^{\dagger} \hat{c}_{i\uparrow}^{\phantom{\dagger}} - \hat{c}_{i\downarrow}^{\dagger} \hat{c}_{i\downarrow}^{\phantom{\dagger}} \rangle$ in the second lead after injecting a spin current polarized in the $z$ direction. 
Both leads have hopping amplitude $t$ in their 2D representation and chemical potential $\mu$. 
There is also Rashba spin-orbit coupling of strength $\lambda$ only in the second lead. 
The coupling to the spin chain $J'$ is tuned to the approximate conducting fixed point, i.e., 
$J'/J = 1.45,1.5,1.7,1.95$ for $\mu/t=0,-1,-2,-3$, respectively. Other parameters are $J/t=2$ and $N_S=12$. 
The measurements are carried out at time $\tau = 70\,t^{-1}$ after a spin voltage $V_S/t = 0.5$ 
is switched on in the first lead at $\tau = 0$. }
\label{figSz}
\end{figure}

\begin{figure}[tb]
\centering
\includegraphics[width=0.96\linewidth]{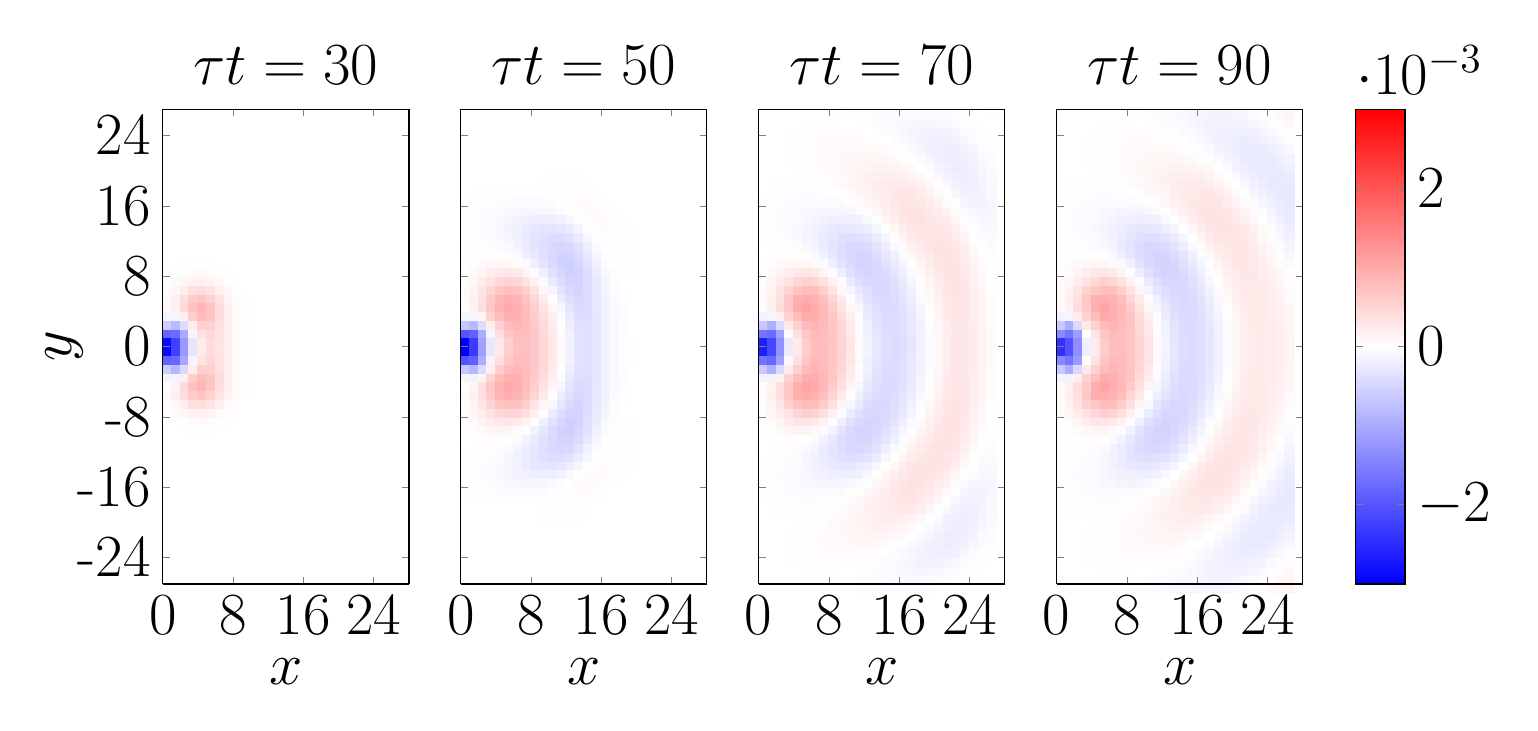}
\caption{(Color online) 
Time-dependence of the magnetization profile for $\lambda/t=0.2$, $\mu/t=-2$ and times $\tau \, t =30,50,70,90$ (from left to right). Other parameters are as in Fig.~\ref{figSz}.
}
\label{figSz2}
\end{figure}

As an example, we investigate the magnetization $m_i = \frac{1}{2} \langle \hat{c}_{i\uparrow}^{\dagger} \hat{c}_{i\uparrow}^{\phantom{\dagger}} - \hat{c}_{i\downarrow}^{\dagger} \hat{c}_{i\downarrow}^{\phantom{\dagger}} \rangle$ in the second lead after a spin current polarized in the $z$ direction is injected. Figure~\ref{figSz} shows $m_i$ for 2D leads with and without spin-orbit coupling. 
The spin current entering the lead induces a small position-dependent magnetization that depends strongly on both the chemical potential $\mu$ and the strength of the spin-orbit coupling $\lambda$. 
Most strikingly, the Rashba precession for $\lambda \neq 0$ leads to oscillations as a function of the position. 
As can be seen in Fig.~\ref{figSz2}, the time-dependence of the local magnetization $m_i$ is relatively small after the wave front has passed, and the Rashba oscillation pattern in particular does not change.

Near half-filling at $\mu=0$, 
the induced magnetization 
is largest along the diagonal directions, 
while it becomes more uniformly spread 
when the chemical potential is decreased. 
This can be understood by looking at the shape of the Fermi surface for $\lambda=0$. 
At half-filling, it takes a diamond form and thus the energy gradient $\nabla_{\bm{k}} E(\bm{k})|_{|\bm{k}| = k_F}$, i.e., the group velocity, points in one of the diagonal directions. 
 In the limit of a nearly empty band $k_F \rightarrow 0$, on the other hand, the Fermi surface becomes a circle and  $\nabla_{\bm{k}} E(\bm{k})|_{|\bm{k}| = k_F}$ is proportional to the momentum $\bm{k}$. 
For finite but small $\lambda$, this picture remains qualitatively valid and the observed angular dependence of the 
magnetization is indeed similar to the case for $\lambda=0$.

Above, it was assumed that the spin current in the spin chain is polarized in the $z$ direction, i.e., orthogonal to the plane of the 2D leads.  
If we choose a different polarization, the spin current through the chain will have the same magnitude because of the pseudo-spin-rotation symmetry of the 1D representation, but the expectation values in the original lattice will differ. 
Instead of carrying out a separate MPS calculation, one could calculate these quantities  
by evaluating formula of Eq.~\eqref{eqExpValBLapprox}  
with the same correlation functions $\langle \hat{a}_{m}^{\dagger} \hat{a}_{n}^{\phantom{\dagger}} \rangle$ and a different transformation matrix $P' = P R$, where $R$ is a unitary matrix that describes the rotation of the pseudo spins.

\section{Conclusions}
\label{secsummary}
We have applied the block-Lanczos DMRG technique to investigate the spin transport in a two-terminal setup consisting of a spin chain and 2D tight-binding leads. 
As long as the spin chain couples only to a single site of each lead, the Lanczos transformation yields an effective 1D model where the leads are semi-infinite chains with nearest-neighbor hopping. 
While the hopping amplitudes are not uniform, their site dependence is negligible except in the vicinity of the chain edge. 
The Lanczos transformation done here can be regarded as a specific case of
the chain mappings for non-interacting baths based on orthogonal polynomials.\cite{BathDiscretizationStrategies} 
Since it is known that these mappings result for typical environments with finite bandwidth in 
asymptotically homogeneous chains,\cite{ChainMappings} 
the effective Hamiltonian we obtained is not surprising and its explicit calculation mostly amounts to determining the 
strength of the inhomogeneities near the spin chain. 
These inhomogeneities can appreciably affect 
the spin transport in the junction because the parameters at the interface need to be fine-tuned to achieve a sizeable spin current at low temperatures.
Qualitatively, however, the behavior of the spin conductance is the same as the case when the hopping strength is assumed to be uniform. 
The Lanczos transformation 
thus does reveal any 
new phenomena in the setup studied here regarding the spin conductance. 
One could apply the method also to more complicated interfaces, e.g. involving multiple coupled spin chains. 
In that case, the lead part would become a ladder model after the transformation, with the number of legs equal to the number of spin chains.  
As realizations of spin chains in solids typically consist of many weakly coupled chains, the block-Lanczos method could be a way towards a more realistic junction model. 

Interestingly, the form of the effective Hamiltonian after the transformation does not change when the Rashba spin-orbit coupling is taken into account. 
Phenomena characteristic of the Rashba model, such as the spin Hall effect,
are hidden in the definition of the Lanczos basis states. 
As a consequence, a spin current entering the lead appears more or less  the same in the 1D representation regardless of the strength of the spin-orbit coupling $\lambda$. 
The inverse Hall effect and the Rashba precession in the original real-space lattice, on the other hand, occur only for finite $\lambda$. By reversing the Lanczos transformation one can calculate quantities in real space and thereby observe these effects.  
However, since the tight-binding leads by themselves are non-interacting, the approach used here only makes sense if the interacting region plays an important role. 
Otherwise, many-body techniques such as the DMRG method are not necessary and more efficient methods, e.g., based on Green's functions, 
are available. 

\section*{Acknowledgments}

DMRG simulations were performed using the ITensor library.\cite{ITensor} 
F. L. was supported by Deutsche Forschungsgemeinschaft through project FE 398/8-1 and 
by the International Program Associate (IPA) program in RIKEN. 
T. S. acknowledges support by Grant-in-Aid for Scientific Research (C) No. 17K05523. S. Y. was supported by 
Grant-in-Aid for Scientific Research (B) No. 18H01183 and in part by JST CREST (Grant Number JPMJCR19J4), Japan.


\end{document}